\documentstyle[11pt]{article}
\bibliography{plain}
\title{\bf Robustness of Entanglement for Bell Decomposable States
} \vspace{20mm}
\author{S. J. Akhtarshenas  $^{a,b,c}$
\thanks{E-mail:akhtarshenas@tabrizu.ac.ir} , M. A. Jafarizadeh$^{a,b,c}$
 \thanks{E-mail:jafarizadeh@tabrizu.ac.ir}
\\
\\
$^a${\small Department of Theoretical Physics and Astrophysics,
Tabriz University, Tabriz 51664, Iran.} \\
$^b${\small Institute for Studies in Theoretical Physics and Mathematics,
Tehran 19395-1795, Iran.} \\
$^c${\small Research Institute for Fundamental Sciences, Tabriz
51664, Iran.}} \pagebreak

\begin{document}
\maketitle \vspace{15mm}
\newpage
\begin{abstract}
We propose a simple geometrical approach for finding the
robustness of entanglement for Bell decomposable states of
$2\otimes 2$ quantum systems. It is shown that the robustness of
entanglement is equal to the concurrence. We also present an
analytical expression for two separable states that wipe out all
entanglement of these states. Finally the random robustness of
these states is also obtained.


{\bf PACs Index: 03.65.Ud }

\end{abstract}
\pagebreak

\vspace{7cm}

\section{Introduction}
During the past decade an increasing study has been made on the
entanglement, although it was discovered several decades ago by
Einstein and  Schr\"{O}dinger \cite{EPR,shcro}. This is because of
the central role that entanglement plays in the theory of quantum
information \cite{ben1,ben2,ben3}. Entanglement as the most non
classical features of quantum mechanics is usually arised from
quantum correlations between separated subsystems which can not be
created by local actions on each subsystem. By definition, a mixed
state $\rho$ of a bipartite system is said to be separable (non
entangled) if it can be written as a convex combination of pure
product states
\begin{equation}
\rho=\sum_{i}p_{i}\left|\phi_{i}^{A}\right>\left<\phi_{i}^{A}\right|
\otimes\left|\psi_{i}^{B}\right>\left<\psi_{i}^{B}\right|,
\end{equation}
where $\left|\phi_{i}^{A}\right>$ and $\left|\psi_{i}^{B}\right>$
are pure states of subsystems $A$ and $B$, respectively. Although,
in the case of pure states of bipartite systems it is easy to
check whether a given state is, or is not entangled, the question
is yet an open problem in the case of mixed states.

There is also an increasing attention in quantifying entanglement,
particularly for mixed states of a bipartite system, and a number
of measures have been proposed \cite{ben3,ved1,ved2,woot}. Among
them the entanglement of formation has more importance, since it
intends to quantify the resources needed to create a given
entangled state.

Also a useful quantities which is introduced in \cite{vidal} as a
measure of entanglement is the robustness of entanglement. It
corresponds to the minimal amount of mixing with separable states
which washes out all entanglement. Analytical expression for pure
states of binary systems have given in \cite{vidal}. Authors in
\cite{du} gave a geometrical interpretation of robustness and
pointed that two corresponding separable states needed to wipe out
all entanglement are necessarily on the boundary of separable set.
Unfortunately, above mentioned quantity as most proposed measures
of entanglement involves extremization  which are difficult to
handle analytically.

In this paper we consider Bell decomposable (BD) states. We
provide a simple geometrical approach and we give an analytic
expression for robustness of entanglement and show that the
corresponding separable states are on the boundary of separable
states as pointed out in \cite{du}. Our approach of determination
of robustness of entanglement is geometrically intuitive. It is
shown that for considered states the robustness is equal to the
concurrence of states. Finally we obtain random robustness for BD
states.

The paper is organized as follows. In section 2 we review BD
states and we present a perspective of their geometry. The
robustness of entanglement of these states is obtained in section
3 via a geometrical approach. Finally the  random robustness is
obtained in section 4. The paper is ended with a brief conclusion.

\section{Bell decomposable states}
In this section we briefly review Bell decomposable (BD) states
and some of their properties. A BD state is defined by:
\begin{equation}
\rho=\sum_{i=1}^{4}p_{i}\left|\psi_i\right>\left<\psi_i\right|,\quad\quad
0\leq p_i\leq 1,\quad \sum_{i=1}^{4}p_i=1,
 \label{BDS1}
\end{equation}
where $\left|\psi_i\right>$ is Bell state, given by:
\begin{eqnarray}
\label{BS1} \left|\psi_1\right>=\left|\phi^{+}\right>
=\frac{1}{\sqrt{2}}(\left|\uparrow\uparrow\right>
+\left|\downarrow\downarrow\right>), \\
\label{BS2}\left|\psi_2\right>=\left|\phi^{-}\right>
=\frac{1}{\sqrt{2}}(\left|\uparrow\uparrow\right>
-\left|\downarrow\downarrow\right>), \\
\label{BS3}\left|\psi_3\right>=\left|\psi^{+}\right>
=\frac{1}{\sqrt{2}}(\left|\uparrow\downarrow\right>
+\left|\downarrow\uparrow\right>), \\
\label{BS4}\left|\psi_4\right>=\left|\psi^{-}\right>
=\frac{1}{\sqrt{2}}(\left|\uparrow\downarrow\right>
-\left|\downarrow\uparrow\right>).
\end{eqnarray}
In terms of Pauli's matrices, $\rho$ can be written as,

\begin{equation}
\rho=\frac{1}{4}(I\otimes I+\sum_{i=1}^{3}
t_i\sigma_{i}\otimes\sigma_{i}), \label{BDS2}
\end{equation}
where

\begin{equation}\label{t-p}
\begin{array}{rl}
t_1=&p_1-p_2+p_3-p_4,  \\
t_2=&-p_1+p_2+p_3-p_4, \\
t_3=&p_1+p_2-p_3-p_4.
\end{array}
\end{equation}

From the positivity of $\rho$ we get
\begin{equation}\label{T1}
\begin{array}{rl}
1+t_1-t_2+t_3\geq & 0,  \\
1-t_1+t_2+t_3\geq & 0,  \\
1+t_1+t_2-t_3\geq & 0,  \\
1-t_1-t_2-t_3\geq & 0.
\end{array}
\end{equation}
These equations form a tetrahedral  with its vertices located at
$(1,-1,1)$, $(-1,1,1)$, $(1,1,-1)$, $(-1,-1,-1)$ \cite{horo2}. In
fact these vertices denote the Bell states  given in Eqs.
(\ref{BS1}) to (\ref{BS4}), respectively.

According to the Peres and Horodecki's condition for separability
\cite{peres,horo1}, a 2-qubit state is separable if and only if
its partial transpose is positive. This implies that $\rho$ given
in Eq. (\ref{BDS2}) is separable if and only if $t_i$ satisfy Eq.
(\ref{T1}) and,
\begin{equation}\label{T2}
\begin{array}{rl}
1+t_1+t_2+t_3\geq & 0,  \\
1-t_1-t_2+t_3\geq & 0,  \\
1+t_1-t_2-t_3\geq & 0,  \\
1-t_1+t_2-t_3\geq & 0.
\end{array}
\end{equation}

Inequalities (\ref{T1}) and (\ref{T2}) form an octahedral with its
vertices located at $O_1^{\pm}=(\pm 1,0,0)$, $O_2^{\pm}=(0,\pm
1,0)$ and $O_3^{\pm}=(0,0,\pm 1)$. So, tetrahedral of Eqs.
(\ref{T1}) is divided into five regions. Central regions, defined
by octahedral, are separable states. There are also four smaller
equivalent tetrahedral corresponding to entangled states. Each
tetrahedral takes one Bell state as one of its vertices. Three
other vertices of each tetrahedral form a triangle which is its
common face with octahedral (See Fig. 1).

\section{Robustness of entanglement}
According to \cite{vidal} for a given entangled state $\rho$ and
 separable state $\rho_{s}$, a new density matrix $\rho(s)$
 can be constructed as,
 \begin{equation}\label{rhos}
 \rho(s)=\frac{1}{s+1}(\rho+s\rho_s),\quad s\geq0,
 \end{equation}
 where it can be either entangled or separable.
 It was pointed that there always exits the minimal $s$
 corresponding to $\rho_s$ such that $\rho(s)$ is separable. This
 minimal s is called the robustness of $\rho$ relative to
 $\rho_s$, denoted by $R(\rho\parallel\rho_s)$. The absolute
 robustness of $\rho$ is defined as the quantity,
 \begin{equation}\label{rob}
 R(\rho\parallel S)\equiv\min_{\rho_s\in S} R(\rho\parallel
 \rho_s).
  \end {equation}

Du et al. in \cite{du} gave a geometrical interpretation of
robustness and pointed that if $s$ in Eq. (\ref{rhos}) is minimal
among all separable states $\rho_s$, i.e. $s$ is the absolute
robustness of $\rho$, then $\rho_s$ and $\rho(s)$ in Eq.
(\ref{rhos}) are necessarily on the boundary of the separable
states.

Here in this section we obtain the absolute robustness for all
Bell diagonal states, and we give an explicit form for the
corresponding $\rho_s$ and $\rho(s)$ which are on the boundary of
the separable states.

Let us consider Fig. 2, we connect $t$, which denotes density
matrix $\rho$, to the center of octahedral such that it cuts the
plane $x_1+x_2+x_3+1=0$ (the boundary of separable octahedral) at
$t^{\prime}$. Then we extend this segment, so that it cuts the
other plane $x_1+x_2+x_3-1=0$ at $t^{\prime\prime}$.

These three points are along the same line but they posses
different lengths. Also it is not difficult to see that they also
lie on planes $x_1+x_2+x_3+\eta=0$, $x_1+x_2+x_3+1=0$ and
$x_1+x_2+x_3-1$, respectively. Using the above argument, we arrive
after some elementary algebra at the following results,
\begin{eqnarray}\label{tp-tpp}
t_i^\prime=\frac{t_i}{\eta}=\frac{-t_i}{t_1+t_2+t_3} \\
t_i^{\prime\prime}=\frac{-t_i}{\eta}=\frac{t_i}{t_1+t_2+t_3}.
\end{eqnarray}

Now using the convexity of the set of density matrices, we can
write $\rho_{s}^{\prime}$ as,
\begin{equation}
\rho^\prime=\frac{1}{1+s}(\rho+s \rho^{\prime\prime}),
\end{equation}
where parameter $s$, called the robustness of $\rho$,  can be
written  as
\begin{equation}
s=\frac{\mid t\,t^{\prime}\mid}{\mid t^{\prime}
\,t^{\prime\prime}\mid}=\frac{1+t_1+t_2+t_3}{2}=C,
\end{equation}
where $C$ is the concurrence of $\rho$ \cite{woot}. The numerical
calculations indicate that thus obtained quantity is minimal with
respect to all separable planes of the octahedral.

In the pioneering paper \cite{vidal}, robustness of entanglement
of Werner states (as a particular kind of BD states), has been
obtained from an entirely different approach. We see that thus
obtained  robustness of entanglement of Werner states via our
proposed procedure is  in agreement with the results of Reference
\cite{vidal}.

 Finally, we would like to emphasis that our treatment is capable to give
an explicit expression  for  the separable matrices
$\rho_s^{\prime}$
 and $\rho_s^{\prime\prime}$.
 Since, using Eqs. (\ref{tp-tpp}) and
(\ref{BDS2}) we can write $\rho_s^{\prime}$ and
$\rho_s^{\prime\prime}$ as
\begin{equation}\label{rhos1}
\rho_s^{\prime}=\frac{1}{4(t_1+t_2+t_3)}\left(
\begin{array}{cccc}
 t_1+t_2 & 0& 0 & -t_1+t_2 \\
  0 &t_1+t_2+2t_3 & -t_1-t_2& 0\\
 0  & -t_1-t_2 & t_1+t_2+2t_3 & 0 \\
 -t_1+t_2 & 0 & 0 & t_1+t_2
 \end{array} \right)
\end{equation}
\vspace{5mm}
\begin{equation}
\rho_s^{\prime\prime}=\frac{1}{4(t_1+t_2+t_3)}\left(
\begin{array}{cccc}
 t_1+t_2+2t_3 & 0& 0 & t_1-t_2\\
  0 & t_1+t_2 & t_1+t_2& 0 \\
  0& t_1+t_2& t_1+t_2 & 0 \\
 t_1-t_2& 0 & 0 & t_1+t_2+2t_3
 \end{array}
\right)
\end{equation}

\section{ Random Robustness}
Also Vidal And Tarrach \cite{vidal} have defined another quantity
so called random robustness, which is defined as robustness of
$\rho$ relative to maximally random state $I/n$. For Bell
decomposable states considered here we can evaluate it as follows.
Using the convexity of the set of density matrices, we can write
$\rho_{s}^{\prime}$ as,
\begin{equation}
\rho^\prime=\frac{1}{1+s_0}(\rho+s_0\,\rho_{0} ),
\end{equation}
where $\rho_{0}=\frac{I}{4}$ and
\begin{equation}
s_0=\frac{\mid t\,t{^\prime}\mid}{\mid
t^{\prime}\,O\mid}=-(1+t_1+t_2+t_3)=2\,C,
\end{equation}
is random robustness of $\rho$. Note that for the states
considered here, the separable matrix $\rho_s^\prime$ has the same
form as given in Eq. (\ref{rhos1}) but with
$\rho_s^{\prime\prime}=\frac{I}{4}$.

\section{Conclusion }
We have  obtained in this work the robustness of entanglement for
Bell decomposable states. It is shown that the corresponding
separable states which  wipe out all entanglement of the states
are on the boundary of separable states. The random robustness of
these states is also obtained.

\newpage

\vspace{10mm}

{\Large {\bf Figure Captions}}

\vspace{10mm}

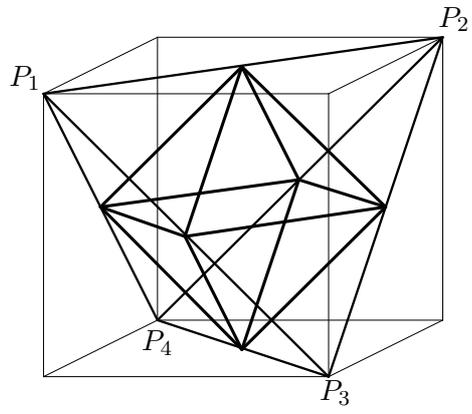
Figure 1: All BD states are defined as points interior to
tetrahedral. Vertices $P_{1}$, $P_{2}$, $P_{3}$ and $P_{4}$ denote
projectors corresponding to Bell states given in Eqs. (\ref{BS1})
to (\ref{BS4}), respectively. Octahedral corresponds to separable
states.

\vspace{10mm}

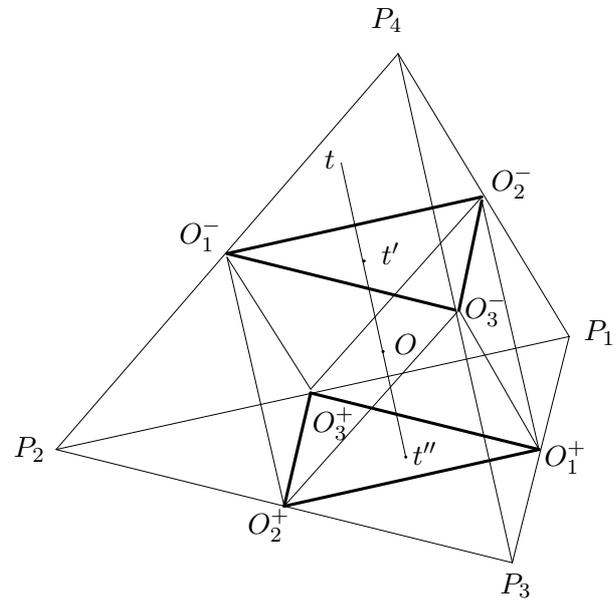
Figure 2: Entangled tetrahedral corresponding to singlet state.
Point $t$ denotes a generic state $\rho$ and points $t^\prime$ and
$t^{\prime\prime}$ are on the separable boundary planes defined by
equations $x_1+x_2+x_3+1=0$ and $x_1+x_2+x_3-1=0$, respectively.


\begin{thebibliography}{99}
\bibitem{EPR}{\sc A. Einstein, B. Podolsky and Rosen, }
{\em  Phys. Rev. {\bf 47}, 777 (1935).}
\bibitem{shcro}{\sc E. Schr\"{O}dinger, }{\em Naturwissenschaften.
 {\bf 23} 807 (1935).}
\bibitem{ben1}{\sc C. H. Bennett, and S. J. Wiesner,}
 {\em Phys. Rev. Lett. {\bf 69}, 2881 (1992).}
 \bibitem{ben2}{\sc C. H. Bennett, G. Brassard,
  C. Cr\'{e}peau, R. jozsa, A Peres and W. K. Wootters,}
 {\em Phys. Rev. Lett. {\bf 70}, 1895 (1993).}
\bibitem{ben3}
{\sc C. H. Bennett, D. P. DiVincenzo, J. A. Smolin and W.K.
Wootters,}
 {\em Phys. Rev. A {\bf 54}, 3824 (1996).}
\bibitem{ved1}{\sc V. Vedral, M. B. Pienio, M. A. Rippin and P. L. Knight}
 {\em Phys. Rev. Lett. {\bf 78}, 2275 (1995).}
\bibitem{ved2}{\sc V. Vedral and M. B. Plenio, }
 {\em Phys. Rev. A {\bf 57}, 1619 (1998).}
\bibitem{woot}{\sc W. K. Wootters, }{\em Phys. Rev. Lett.
{\bf 80} 2245 (1998).}
\bibitem{vidal}{\sc G. Vidal and R. Tarrach, }
{\em  Phys. Rev. A {\bf 59,} 141 (1999).}
\bibitem{du}{\sc J.F. Du, M.J. Shi, X.Y. Zhou and R.D. Han, }
{\em  Phys. Lett. A {\bf 267,} 244 (2000).}
\bibitem{peres}{\sc A. Peres, }
{\em Phys. Rev. Lett. {\bf 77} 1413 (1996).}
\bibitem{horo1}{\sc M. Horodecki, P. Horodecki and R. Horodecki, }
{\em Phys. Lett. A  {\bf 223} 1 (1996).}
\bibitem{horo2}{\sc R. Horodecki and M. Horodecki }
{\em Phys. Rev. A {\bf 54} 1838 (1996).}

\end{thebibliography}
\end{document}